\documentclass{elsart}

\usepackage{amsmath}
\usepackage{amssymb}
\usepackage{graphicx}
\usepackage{dcolumn}
\usepackage{bm}
\usepackage{color}
\usepackage{array}
\newcommand{\PreserveBackslash}[1]{\let\temp=\\#1\let\\=\temp}
\newcolumntype{C}[1]{>{\PreserveBackslash\centering}p{#1}}
\newcolumntype{R}[1]{>{\PreserveBackslash\raggedleft}p{#1}}
\newcolumntype{L}[1]{>{\PreserveBackslash\raggedright}p{#1}}

\begin{document}

\begin{frontmatter}

\title{Quantum Barro-Gordon Game in Monetary Economics}

\author[1]{Ali Hussein Samadi},
\author[2]{Afshin Montakhab}\ead{montakhab@shirazu.ac.ir},
\author[1]{Hussein Marzban},
\author[1]{Sakine Owjimehr}

\address[1]{Department of Economics, College of Economics Management and Social Sciences, Shiraz University, Shiraz 71946-85111, Iran}
\address[2]{Department of Physics, College of Sciences, Shiraz University, Shiraz 71946-84795, Iran}

\begin{abstract}

Classical game theory addresses decision problems in multi-agent
environment where one rational agent's decision affects other
agents' payoffs. Game theory has widespread application in
economic, social and biological sciences. In recent years quantum
versions of classical games have been proposed and studied. In
this paper, we consider a quantum version of the classical
Barro-Gordon game which captures the problem of time inconsistency
in monetary economics. Such time inconsistency refers to the
temptation of weak policy maker to implement high inflation when
the public expects low inflation. The inconsistency arises when
the public punishes the weak policy maker in the next cycle. We
first present a quantum version of the Barro-Gordon game. Next, we
show that in a particular case of the quantum game,
time-consistent Nash equilibrium could be achieved when public
expects low inflation, thus resolving the game.

\end{abstract}

\begin{keyword}
Barro-Gordon Game\sep Quantum Game Theory\sep Time Inconsistency

\PACS 03.65 Aa \sep 03.65.Ud\sep 03.65.-w\sep01.80.+b\sep 02.50.Le\\

HIGHLIGHTS\\
$\bullet$•    We reformulate Barro-Gordon Game using  quantum game theory.\\
$\bullet$•    We use Marinatto-Weber approach for quantization of game theory.\\
$\bullet$•    We find that the well-known time inconsistency in
the classical game is removed after quantization.
\end{keyword}

\end{frontmatter}

\section{Introduction}

Game theory is a mathematical formulation of  situations where,
for two or more agents, the outcome of an action by one of them
depends not only on the particular action taken by that agent but
on the actions taken by the other (or others)\cite{Carmichael
2005}. Game theory has gained much attention in recent years as
indicated by the many texts written on it \cite{Gibbons 1992, Borm
and peters 2002, Vega-Redondo 2003, ken2007}. After Neuman and
Morgenstern's book \cite{neuman}, which was the first important
text in game theory, John Nash made important contributions to
this theory \cite{nash1, nash2}. However, application  of game
theory to different areas such as economics, politics, biology,
started in 1970's and has been growing ever since \cite{Carmichael
2005}.

In recent years, game theory has attracted the attention of many
physicists as well. Among the many contributions that has come
along from the physics community is the inclusion of the rules of
quantum mechanics in game theory where quantum effects such as
superposition and entanglement can play a role
\cite{griffiths2005, dirac 2012}. Quantum game was introduced by
Eisert, Wilkens, and Lewenstein (EWL) \cite{eisert 1999}, where
the role of entanglement was considered first.  Subsequently,
Marinatto and Weber (MW) \cite{mw2000} offered a more general
scheme for quantization of games based on Hilbert space approach.
It is now believed that quantization of games can offer
interesting situations where various dilemma present in classical
games can be removed. As a result of the above-mentioned works,
many authors have employed the quantization of various games. For
example, Cheon and Tsutsui \cite{cheon 2006}, Fliteny and
Hollenberg \cite{fliteny 2007}, Makowski \cite{makowski 2009} and
Landsburg \cite{landsburg 2011} have used EWL approach.  On the
other hand, Arfi \cite{arfi 2007}, Deng et. al \cite{deng 2016}
and Frackiewicz \cite{frackiewicz 2015}, based their approach on
MW scheme. Much of the attention in the above-mentioned works has
been paid to well-known games such as prisoner's dilemma game
\cite{eisert 1999,cheon 2006,fliteny 2007,arfi 2007}, while others
have considered various other games with social implications.

However, despite the growing interest in econophysics
\cite{MS2000}, little attention has been paid to quantization of
classical games in finance and economics. The present work offers
a step in this direction. Here, we intend to quantize the
classical game of Barro and Gordon (BG) in monetary economics
employing the MW scheme.  We then study some specific cases of the
quantum game in order to find Nash equilibrium and the advantages
quantization may offer. BG game, to be explained in details in the
following, is a classical game which illustrates the problem of
time inconsistency in monetary policy. Time inconsistency was
first introduced by Kydland and Prescott \cite{kydland 1977} and
later by Barro and Gordon \cite{bg 1983}. The main idea is that
when the output is inefficiently low, policy maker can increase it
by applying discretionary policy, causing surprising inflation. In
this situation, although the output increases which is beneficial,
it causes inflation, which is costly. Therefore, we encounter an
inflationary bias. Since BG introduced a noncooperative game
between public and central bank, many researchers \cite{lucas,
albanesi, king, demirel,adam} have been using game theory to study
time inconsistency. All of these studies are based on the
classical game theory, while the principles of quantization has
not been applied so far.

There are, however, a variety of reasons to apply the roles of
quantum mechanics to various disciplines outside of traditional
physics. For example, in psychology and decision making theory,
quantum cognition has gained much attention \cite{ bruza,
Bordley}. In most such approaches, quantum probabilities are
considered in order to represent certain uncertainty in decision
making process. Furthermore, a quantum Hamiltonian approach to
various fields of social sciences has also been considered \cite{
Bleh, Bagarello, Bagarello1, Haven, Bagarello2, Bagarello3} where
operator-valued dynamical variables are governed by a general
Hamiltonian which includes all possible interaction. However, we
must note that while our approach in this paper is similar in
spirit, it is different in its basic assumption and methodology.
The problem of strategy selection which is the essence of game
theory could be formulated using the laws of quantum mechanics
instead of classical probability theory. In fact, it is the purely
quantum mechanical concept of superposition (of strategies) which
provides the key ingredient in our approach to game theory here.

Furthermore, micro-evolution teaches us that our selfish
(microscopic) genes may make our decision on a fundamental level,
where quantum mechanics may be relevant.  Perhaps from a more
practical point of view, recent advances in quantum computational
and quantum communication technology\cite{ni} may help us in
creating quantum devices which must take on quantum strategies in
order to solve problems \cite{eisert 1999}. It is with such
motivations in mind that one may consider quantum game theory in
various fields of social sciences, and consequently monetary
economics in our present case.

The rest of this paper is organized as follows:  In Section 2, we
briefly describe the problem of time inconsistency in monetary
policy. The classical BG game is then described in details in
Section 3. In Section 4, we present our main results of
quantization of BG game and consider some specific cases of the
quantum game.  The last section offers concluding remarks.

\section{Time Inconsistency}

A policy is time inconsistent or dynamic inconsistent when it is
considered as the best policy for particular time in the future,
but it does not remain so, when that particular time actually
arrives. There are two possible mechanisms that have been
considered for such a time inconsistency. (i) Strotz \cite{strotz
1955} explains that time inconsistency is because of changing
preferences, (ii) Kydland and Prescott \cite{kydland 1977}
consider another explanation that is based on agent's rational
expectation. The main idea is, when people expect low inflation,
central bank finds the incentive for high inflation. If the public
understand this incentive and predict high inflation, the central
bank finds it optimal to deliver the public's expectation and
therefore implements high inflationary policy. Therefore, while
low inflation is the optimal policy for both banker and the public
to begin with, high inflation is eventually implemented. We use
this second mechanism in this paper as it is the concept of time
inconsistency associated with the Barro-Gordon game.

Barro and Gordon \cite{bg 1983} have explained time inconsistency
of monetary policy as follow: in a discretionary regime, central
banker can print more money and make more inflation than people's
expectations. Benefits of this surprising inflation might provide
more economic activities or reduce government's debt. However, if
people, due to their rational expectation, understand it and
adjust their expectations with it, then policy maker will not
reach his goal at all. This is simply due to the fact that
inflationary advantages is best when it is unanticipated. Besides
that, due to increased money supply, the level of prices will
grow, which will have negative consequences for the policy maker.
The classical Barro-Gordon game captures the essence of this type
of time-inconsistency in the context of decisions (strategy) that
a policy maker must make and the expectations that the public can
have. The actual game is represented in two different formats
where the strong policy maker implements low inflation which is
time consistent, while in the case of weak policy maker a time
inconsistent strategy is the alternative to a Nash equilibrium.

\section{BG  Classical Game   }

In BG game, similar to prisoners's dilemma, there are two players;
public and central bank. In this game, one assumes that the public
has rational expectations. The public then predicts inflation by
solving out the policy maker's optimization problem. On the other
hand, the policy maker selects inflation policy by considering the
public's inflationary expectations. In this paper, by following
Backus and Driffill \cite{backus 1985} and Storger \cite{storger
2007}, we use a special version of BG game. This version is easier
to convert to quantum game due to having a definitive payoff
matrix. In this version, there are two types of policy maker: weak
policy maker, which uses discretionary policies and gains benefit
by making unanticipated inflation. In the other words, a weak
policy maker can cheat the public when they formed their low
inflationary expectation at the start of the period. Strong policy
maker, on the other hand,  commits to zero inflation and is not
interested in unanticipated inflation.

The utility function of these policy makers is as follows:
\begin{equation}
U^{pol}_{t}=\theta b (\pi_t\ - \pi^{e}_ {t}\ )-\frac{a \pi^{2}_{t}}{2} \label{Sim1}
\end{equation}
where, $\pi_t\ $ and  $\pi^{e}_{t}\ $are actual inflation and
expected inflation rates, respectively. Inflationary cost is
assumed to be proportional to the square of inflation and
therefore $ \frac{a \pi^{2}_{t}}{2}$ is the cost of inflation
where $a$ is an arbitrary cost parameter. $\theta\ $ is a dummy
variable that is equal to 1 for weak policy maker and 0 for strong
policy maker and $b$ is a coefficient for benefit of inflation
term with $b\
>0\ $. If $\pi_t\ > \pi^{e}_{t}\ $then policy maker can decrease
unemployment (according to Philips curve \footnote[1]{Philips
curve shows an inverse relation between the unemployment and
inflation.}) and gain benefit using the first term in Eq.(1).
Public's utility function is as follows:
 \begin{equation}
 U^{pub}_{t}=-(\pi_t\ - \pi^{e}_ {t}\ )^2. \label{Sim2}
 \end{equation}
This function shows that every deviation from expected inflation
causes disutility for  the public.

We next briefly review the payoff matrix for weak and strong
policy maker as obtained by \cite{backus 1985,storger 2007}: First
we use weak policy maker optimization. The weak policy maker
optimizes Eq.(1) without any constraint. By taking derivative of
Eq.(1) with respect to $\pi_t$ , optimal inflation will be $ \hat
\pi_t = \frac {b} {a}$. If  $ \pi^{e}_{t} = \frac {b} {a}$,
replacing it in Eq.(1) and (2) will result in $ U^{pol}_{t}= -
\frac {b^2}{2a}<U^{pub}_{t}=0 $. Therefore, in this case, weak
policy maker cannot gain any benefit and this strategy will not be
chosen. On the other hand, optimizing unconstraint Eq.(2), with
the assumption of public rational expectations, results in $\pi_t
= \pi^{e}_ {t}$. If weak policy maker commits to zero inflation,
both players will receive zero payoff. However, if the public
expects zero inflation, and the weak policy maker implements $
\pi_t = \frac {b} {a}$, then he can gain some benefit (equal to $
\frac {b^2} {2a}$) and the public will incur losses of $ -( \frac
{b} {a})^2 $ . Therefore, we have $ U^{pol}_{t}=  \frac
{b^2}{2a}>U^{pub}_{t}=-(\frac {b}{b})^2 $  and the weak policy
maker therefore prefers this strategy. Even if the public expects
$ \pi^{e}_{t} = \frac {b} {a}$  and the policy maker chooses $
\pi_t = \frac {b} {a}$, he can get more payoff than choosing zero
inflation. Thus, $ \pi_t = 0$ is a dominated strategy and will
never be selected. Following \cite{backus 1985}, normalization
condition ($ a=b=2 $), leads to a simple payoff matrix for the
weak policy maker as shown in Table 1. Note that the case of $
\pi_t = 1$ , $ \pi^{e}_{t} = 1$ is a Nash equilibrium in this
case. However, the actual equilibrium is the case of $ \pi_t = 1$
, $ \pi^{e}_{t} = 0$ \emph{if} the policy maker is successful in
cheating the public . The key point here is that this equilibrium
($ \pi_t = 1$ , $ \pi^{e}_{t} = 0$) is time inconsistent because $
\pi^{e}_{t} = 0$ is announced but $ \pi^{e}_{t} = 1$ is
implemented.

\begin{table}[h!]
\centering
\begin{center}
\begin{tabular} {c|ccc}

  \hline \hline
   $\theta =1$     &    &  Public  &    \\
  \hline
     &                &  $\pi^{e}_{t} = 0$  &  $\pi^{e}_{t} = 1$ \\
   Weak policy maker  & $\pi_t = 0$  &  (0,0)  &  (-2,-1)   \\
     &                 $\pi_t = 1$  &  (1,-1)  &  (-1,0)    \\
   \hline \hline

\end{tabular}
\caption{Weak policy maker payoff} \label{weak policy maker
payoff}
\end{center}
\end{table}

In the case of strong policy maker ($\theta=0\ $), $ \pi_t = \frac
{b} {a}$ is never chosen as it is a dominated strategy.  Strong
policy maker will incur a  loss equal to$  \frac {-b^2} {2a}$ in
both cases (either the public expects zero inflation or $\frac {b}
{a}$). Therefore, $ \pi_t = 0$ is the best policy for strong
policy maker and therefore always commits to it. In this
situation, there would be no problem of time inconsistency.
Therefore, ($ \pi_t = 0$, $ \pi^{e}_{t} = 0$) is the Nash
equilibrium for this case and it is time consistent.
\begin{table}[h!]
\centering
\begin{center}
\begin{tabular} {c|ccc}

  \hline \hline
   $\theta =0$     &    &  Public  &    \\
  \hline
     &                &  $\pi^{e}_{t} = 0$  &  $\pi^{e}_{t} = 1$ \\
   Strong policy maker  & $\pi_t = 0$  &  (0,0)  &  (0,-1)   \\
     &                 $\pi_t = 1$  &  (-1,-1)  &  (-1,0)    \\
   \hline \hline

\end{tabular}
\caption{Strong policy maker payoff} \label{strong policy maker
payoff}
\end{center}
\end{table}

Barro and Gordon \cite{bg 1983} showed that the weak policy maker
loses his reputation for cheating the public. In fact, in the next
period, public plays ``tit for tat" game and punish the weak
policy maker by adjusting their expectations. In other words, if $
\pi_{t-1} =  \pi^{e}_{t-1} $then $ \pi_t =  \pi^{e}_{t} = 0$ ;
otherwise, $ \pi_t =  \pi^{e}_{t} = \frac {b} {a}$. Therefore,
weak policy maker compares marginal cost and benefit of cheating
the public, and then decides to make unanticipated inflation.
Consequently, classical BG game needs two time periods to solve
the game between the public and weak policy maker. In this paper
we generalize this classical game to a quantum framework and ask
if it can be made more efficient.

\section{Quantum  BG  Game}
\subsection{ Quantization of the game} \

There is essentially two different ways to quantized classical
games in the literature. EWL \cite{cheon 2006,fliteny
2007,makowski 2009,landsburg 2011} took the original steps in this
regard. However, the approach of MW \cite{arfi 2007, deng 2016,
frackiewicz 2015} has been more widely used recently and we intend
to use their approach in quantizing BG game.

Fr¸ackiewicz \cite{frackiewicz 2015} argued that, in EWL method,
the result of the game depends on many parameters because each
player's strategy is a unitary operator. Therefore it has
cumbersome calculation. But in MW, player's local operators were
performed on some fixed entangled state $\left| \psi\right \rangle
$. It seems that MW is simpler than EWL \cite{epjb2016}. We
therefore use the MW method to quantize the game between weak
policy maker and the public. In this game, there are two players:
weak policy maker (M) and public (U). Each player has two
strategies: high inflation (H) and low inflation (L). Consider a
four- dimensional Hilbert space, $ \mathcal{H} \ $, as the
strategy space in ket notion:
\begin{equation}
\mathcal{H}=\mathcal{H}_M \otimes \mathcal{H}_U=\left \{ \left|LL \right \rangle,\left|LH \right \rangle,\left|HL \right \rangle ,\left|HH \right \rangle \right \}  \label{Sim3}
\end{equation}
where the first qubit is related to the state of the policy maker
and the second one to that of public. Kets show a  given strategy
in strategy space which in quantum version is a Hilbert space.
Therefore, we use an arbitrary quantum strategy as a normalized
state vector,$\left| \psi_{i}\right \rangle $.
\begin{equation}
\left| \psi_{i}\right \rangle =\alpha\left|LL \right \rangle+\gamma\left|LH \right \rangle+\delta\left|HL \right \rangle +\beta\left|HH \right \rangle \
\end{equation}
Where $|\alpha|^2 , |\beta|^2 , |\gamma|^2 , |\delta|^2 $ are
probability of observing the strategies of (L, L), (H, H), (L, H)
and (H, L), respectively, with  $
|\alpha|^2+|\gamma|^2+|\delta|^2+|\beta|^2 =1 $. Density matrix is
written as $ \rho_{i}= \left| \psi_{i} \right \rangle\left \langle
\psi_{i} \right | $. Let C be a unitary Hermitian operator (i.e, $
C^\dagger=C=C^{-1}$), such that $C\left|H \right \rangle=\left|L
\right \rangle $ and $C\left|L \right \rangle=\left|H \right
\rangle $ and  $ \mathcal{I} \ $ is the identity operator such
that $ \mathcal{I}\left|H \right \rangle=\left|H \right \rangle $
and $\mathcal{I} \left|L \right \rangle=\left|L \right \rangle $ .
In the game, policy maker and [public] use operators $ \mathcal{I}
\ $ and C with probabilities of $ p,(1-p),[q,(1-q)]$.

Final density matrix for this system is as follows:

  \begin{eqnarray}
    \rho_{f}
   =&&pq\left [ (\mathcal{I}_M \otimes \mathcal{I}_U)\rho_{i}(\mathcal{I}^\dagger_M \otimes \mathcal{I}^\dagger_U) \right ] \nonumber \\
   &&+p(1-q)\left [ (\mathcal{I}_M \otimes C_U)\rho_{i}(\mathcal{I}^\dagger_M \otimes C^\dagger_U) \right ] \nonumber
   \\
   &&+(1-p)q\left [ (C_M \otimes \mathcal{I}_U)\rho_{i}(C^\dagger_M \otimes \mathcal{I}^\dagger_U) \right ] \nonumber \\
   && + (1-p)(1-q) \left [ (C_M \otimes C_U)\rho_{i}(C^\dagger_M \otimes C^\dagger_U) \right ]
   \end{eqnarray}
   and the two payoff operators are given as follows:
   \begin{equation}
   P_M=0\left|LL \right \rangle\left\langle\ LL \right| -2\left|LH \right \rangle\left\langle\ LH \right|+\left|HL \right \rangle\left\langle\ HL \right| -\left|HH \right \rangle \left\langle\ HH \right|
   \end{equation}                                                     \begin{equation}
      P_U=0\left|LL \right \rangle\left\langle\ LL \right| -\left|LH \right \rangle\left\langle\ LH \right|-\left|HL \right \rangle\left\langle\ HL \right| +0\left|HH \right \rangle \left\langle\ HH \right|.
      \end{equation}
Finally, payoff functions are calculated according to:
\begin{equation}
\bar \$_M(p,q)=Tr(P_M \rho_{f})
\end{equation}
\begin{equation}
\bar \$_U(p,q)=Tr(P_U \rho_{f}).
\end{equation}
This can be written as:
\begin{equation}
\bar \$_M(p,q)=\boldsymbol{\Phi} \boldsymbol{\Omega} \gamma^T_M
\end{equation}
\begin{equation}
\bar \$_U(p,q)=\boldsymbol{\Phi} \boldsymbol{\Omega} \gamma^T_U
\end{equation}
where
\begin{equation}
\boldsymbol{\Phi}=\left [pq , p(1-q), (1-p)q, (1-p)(1-q)\right ]
\end{equation}

\begin{center}
 $\boldsymbol{\Omega}$ =
$\begin{bmatrix}
 \alpha^2 & \gamma^2 & \delta^2 & \beta^2 \\
 \delta^2 & \beta^2  & \alpha^2 & \gamma^2 \\
 \gamma^2 & \alpha^2 & \beta^2  & \delta^2  \\
 \beta^2  & \delta^2 & \gamma^2 & \alpha^2
\end{bmatrix}$ \\
$ \gamma_M=\left [ 0 , -2, 1, -1\right ] $ \\
$ \gamma_U=\left [ 0 , -1, -1, 0\right ] $. \\
\end{center}

 The payoff functions for policy maker and public are therefore calculated as:
\begin{equation}
\bar \$_M(p,q)=2p(\alpha^2-\beta^2+\delta^2-\gamma^2)+ q(\delta^2-\alpha^2-\gamma^2+\beta^2)-\alpha^2+\gamma^2-2\delta^2
\end{equation}
\begin{equation}
\bar \$_U(p,q)=(1-2(\delta^2+\gamma^2)) (q(2p-1)-p)-(\delta^2 +\gamma^2)
\end{equation}
In order for Nash equilibrium to exist one needs to implement  the following conditions \cite{mw2000}:
\begin{equation}
\bar \$_M(p^*,q^*)- \bar \$_M(p,q^*)\ge 0, \qquad  \forall p\in[0,1]\\
\end{equation}
\begin{equation}
\bar \$_U(p^*,q^*)-\bar \$_U(p^*,q)\ge 0,  \qquad  \forall q\in[0,1]\\
\end{equation}
which in our case lead to:
\begin{equation}
2(p^*-p)(\alpha^2-\beta^2+\delta^2-\gamma^2)\ge 0
\end{equation}
\begin{equation}
(1-2(\delta^2+\gamma^2))(q^*-q)(2p^*-1)\ge 0.
\end{equation}
\subsection{Analysis of the game}

Equation (13-14) and (17-18) are our main results. Following MW's
approach, we consider the validity of three possible situations
below:

(a) $ p^*=q^*=1 $

In this case the payoffs are as follows:
\begin{equation}
\bar \$_M(1,1)=-\beta^2-2\gamma^2+\delta^2
\end{equation}
\begin{equation}
\bar \$_U(1,1)=-\gamma^2-\delta^2
\end{equation}
And Nash equilibrium conditions are:
\begin{equation}
2(1-p)(\alpha^2-\beta^2+\delta^2-\gamma^2)\ge 0
\end{equation}
\begin{equation}
(1-2(\delta^2+\gamma^2))(1-q)\ge 0\Rightarrow \gamma^2+\delta^2 \le 1/2.
\end{equation}
If $\alpha^2+\delta^2>\beta^2+\gamma^2$ , then the first Nash
equilibrium condition will be satisfied. In fact, this condition
will most likely be satisfied for the $ \mathit{weak} $ policy
maker which we are considering here, since according to Table (1),
he prefers (L,L) or (H,L) strategies rather than (H,H) or (L,H)
strategies. We thus assume that for weak policy maker the
condition of   $\alpha^2+\delta^2>\beta^2+\gamma^2$ is always
true. The second Nash equilibrium condition may or may not be
satisfied depending on the choice of the quantum strategy. We will
return to this point later on in this paper.

(b) $ p^*=q^*=0 $

In this case the payoffs are as follows:
\begin{equation}
\bar \$_M(0,0)=-\alpha^2+\gamma^2-2\delta^2
\end{equation}
\begin{equation}
\bar \$_U(0,0)=-\gamma^2-\delta^2
\end{equation}
and Nash equilibrium conditions are calculated as:
\begin{equation}
-2p(\alpha^2-\beta^2+\delta^2-\gamma^2)\ge 0
\end{equation}
\begin{equation}
(1-2(\delta^2+\gamma^2))q\ge 0\Rightarrow \gamma^2+\delta^2 \le 1/2
\end{equation}

In this case, Nash equilibrium will be satisfied if
$\alpha^2+\delta^2<\beta^2+\gamma^2$  which is exactly the
opposite of the previous case. Again, since we are considering a
weak policy maker here (see above), we will consider this
condition as unacceptable. Therefore, we do not consider Nash
equilibrium to hold for the weak policy maker in the case of
$p^*=q^*=0$.

(c) $ p^*=q^*=1/2 $

In this case the payoffs are as follows:
\begin{equation}
\bar \$_M(1/2,1/2)=-1/2
\end{equation}
\begin{equation}
\bar \$_U(1/2,1/2)=-1/2
\end{equation}
and Nash equilibrium conditions are calculated as:
\begin{equation}
2(1/2-p)(\alpha^2-\beta^2+\delta^2-\gamma^2)\ge 0
\end{equation}
\begin{equation}
(1-2(\delta^2+\gamma^2))(1/2-q)(1-1)=0
\end{equation}
The second Nash equilibrium condition is trivially satisfied.
However, the first Nash equilibrium condition is clearly violated
for $p>1/2$ for a weak policy maker (i.e.
$\alpha^2-\beta^2+\delta^2-\gamma^2 >0$). But since in this
scenario both players have negative payoff, this equilibrium is
not preferred.

Therefore among the three possibilities we have considered above,
the second scenario (b) could not be satisfied because Nash
equilibrium did not exist and the third scenario (c) included
dominated strategies. We therefore choose to only consider the
first scenario (a). However, the first scenario could satisfy Nash
equilibrium depending on the choice of quantum strategies, i.e.
choice of $\alpha,\beta,\delta,\gamma $ . In the following we
consider two different quantum strategies of weak policy maker
where Nash equilibrium could potentially exist:

(i) Suppose that the public has false prediction about
inflation. It means that, quantum strategy is a superposition of
two strategies, (L, H) and (H, L) where $\alpha=\beta=0$:
\begin{equation}
\left| \psi_{i}\right \rangle =\gamma\left|LH \right \rangle+\delta\left|HL \right \rangle . \
\end{equation}
Therefore, payoff functions will be as follows:
\begin{equation}
\bar \$_M(1,1)=-2\gamma^2+\delta^2
\end{equation}
\begin{equation}
\bar \$_U(1,1)=-\gamma^2-\delta^2=-1
\end{equation}
And Nash equilibrium conditions:
\begin{equation}
2(1-p)(\delta^2-\gamma^2)\ge 0
\end{equation}
\begin{equation}
 \gamma^2+\delta^2 \le 1/2
\end{equation}

In this case, the public would always lose due to false
expectations, i.e. Eq. (33). However weak policy maker can earn
better a payoff if $ \delta^2> 2\gamma^2$. However, this scenario
cannot imply a stable situation due to Eq. (35) which indicate
that Nash equilibrium can never be obtained since
$\gamma^2+\delta^2=1$. This result is reminiscent of the classical
version of the game where the weak policy maker can earn positive
payoff by cheating the public for just one cycle. Afterwards, the
public will punish him and correct their expectation. Here, we
showed  that a quantum strategy that is superposition of false
prediction strategies is not a Nash equilibrium. In other word, it
is not  a sustainable equilibrium.

(ii) Suppose that public has correct prediction about the inflation.
Therefore, quantum strategy is a superposition of two strategies, (L, L) and (H, H) where $\gamma=\delta=0$:
\begin{equation}
\left| \psi_{i}\right \rangle =\alpha\left|LL \right \rangle+\beta\left|HH \right \rangle  \
\end{equation}
\begin{equation}
\bar \$_M(1,1)=-\beta^2
\end{equation}
\begin{equation}
\bar \$_U(1,1)=0
\end{equation}
Thus Nash equilibrium conditions in this case are:
\begin{equation}
2(1-p)(\alpha^2-\beta^2)\ge 0
\end{equation}
\begin{equation}
(1-q)\ge 0
\end{equation}
The second Nash equilibrium condition (Eq. 40) is always true.
However, the first one (Eq. 39) will be satisfied for
$\alpha^2>\beta^2$ which is an acceptable condition for the weak
policy maker. This shows that the larger the share of (L, L)
strategy is chosen ($\alpha$) the smaller the negative payoff of
policy maker becomes ($\beta$). The important point here is that a
Nash equilibrium exist for the weak policy maker where the public
is guaranteed not to lose and the policy maker's loss can be
minimized by reducing $\beta $.  In fact, in an extreme case
$\beta \rightarrow 0$, where the quantum strategy converges to a
(non-superposition) single strategy (L,L), the payoff of both
players will be zero as in the classical case (see Table (1)).
However the important difference is that Nash equilibrium is
satisfied here, where in the classical version it is not.
Therefore, the quantum version of the game in this scenario offers
a time consistent Nash equilibrium.  This constitutes our main
result.

\section{Concluding Remarks}

Barro and Gordon proposed a game between the public and policy
maker based on the theory of time inconsistency. In this game, a
weak policy maker can earn some benefits in short time by cheating
the public about inflation. However, the public will punish him in
the next period. Therefore, inflation increases and policy maker
will lose his benefits. Thus, in this classical version of BG
game, the implementation of low inflation by the weak policy maker
is not a Nash equilibrium. In this paper we generalized the BG
game by using the quantum game scheme according to Marinatto and
Weber. We considered the quantum game as a superposition of four
classical strategies, and Nash equilibrium conditions were
subsequently calculated. The results showed that among the three
possibilities we have considered, the first scenario was more
acceptable. Then we considered two different quantum strategies of
weak policy maker where Nash equilibrium could potentially
exist:(i) public has false prediction and (ii) public has a
correct prediction. It was shown that Nash equilibrium was not
satisfied when the public has false prediction. However, we
obtained a Nash equilibrium that is time consistent in the second
scenario where the public has a correct prediction about
inflation. Our result is important since it shows that in the
quantum version of BG game, unlike its classical version, the
\emph{low inflation} policy is a Nash equilibrium when the public
expects low inflation thus removing the time inconsistency and
therefore solving the game.  We emphasize that the purely quantum
effect of superposition entangled states was the key ingredient in
solving the game and removing the time inconsistency present in
the classical version.

We next briefly comment on some issues regarding our results. The
relevance of quantum game may at first glance seem a bit peculiar
despite the motivations provided in the Introduction (Section 1)
above.  Ever since 1935 when Schrodinger introduced what
is now known as the Schrodinger cat, the possibility of
macroscopic superposition states was debated in the literature. However,
quantum technology has provided for macroscopic superposition
states \cite{nature2000}, and
one can imagine that with sufficiently advanced technology, future machines
could employ strategies that could benefit from quantum game theory.  Furthermore,
one may ask if our results would be different if we had employed other methods
besides MW for game quantization. Arfi\cite{arfi 2007} has shown that one would
obtain the same results for prisoner's dilemma game regardless of the method of
quantization.  We chose the MW method since it suited our game in a more straightforward
way. However, one can imagine that employing the method of EWL would lead to
essentially the same results.  The important point that seems to be the common point of
most game quantization is that quantum games offer an advantage over their classical
version because they employ superposition principle and are thus able to resolve
the conflict existing in the classical version.  We have also obtained the same essential
results here, and suspect that our result would be independent of the method of quantization.

Our aim here has been to provide an example of quantum game theory in economics and
how the rules of quantum mechanics may offer advantages in this regard.  However, one
might consider further work along the same line presented here.  For example, one can
consider the possibility of other equilibria that might exist for the case of
various other choices of $p$ and $q$ besides the ones considered here (which
were purely motivated by previous studies).  Another interesting avenue would be
to consider a Hamiltonian formulation and thus the time evolution of various operators
along the line of \cite{Bagarello,Bagarello2,Bagarello3}.  This might be interesting as dynamical
evolution would become quantum mechanical and one might consider the different evolution
of an initially (quantum) superposition state vs. its classical analog of a mixed state.

\emph{Acknowledgements--}Grants from Research Council of Shiraz
University is kindly acknowledged. This paper has also benefited from
constructive criticism of respected (anonymous) referees.


\begin{thebibliography}{0}

\bibitem{Carmichael 2005}F. Carmichael,\textit{ A Guide to Game Theory}, Financial Times Prentice Hall, 2005.
\bibitem{Gibbons 1992}R. Gibbons, \textit{Theory for Applied Economists}, Princeton University Press, 1992.
\bibitem{Borm and peters 2002} P. Borm  H. J. Peters, \textit{Chapters in Game Theory In honor of Stef Tijs}, Kluwer Academic Publishers, 2002.
\bibitem{Vega-Redondo 2003}F. Vega-Redondo,\textit{ Economics and the Theory of Games}, Cambridge University Press, 2003.
\bibitem{ken2007} K. Binmore, Playing for Real, \textit{A Text on Game Theory}, Oxford University Press, 2007.
\bibitem{neuman} J. Von Neuman and  O. Morgenstern, \textit{Theory of Games and Economic Behaviour}. NY: Wiley, 1944 (Reprint 1964).
\bibitem{nash1} J. F. Nash, \textit{The Bargaining Problem}, Econometrica. 18 (1950) 155.
\bibitem{nash2} J. F. Nash, \textit{Two-Person Cooperative Games}, Econometrica. 21 (1953) 128.
\bibitem{griffiths2005} D. J. Griffiths,\textit{ Introduction to Quantum
Mechanics}, Second Edition, Upper Saddle River, NJ: Pearson, 2005.
\bibitem{dirac 2012}P. A. M. Dirac, \textit{Principles of Quantum Mechanics}, Oxford University Press, 2012.
\bibitem{eisert 1999} J. Eisert, M. Wilkens, M. Lewenstein, \textit{Quantum Games and Quantum Strategies}, Physical Review Letters 83 (1999) 3077.
\bibitem{mw2000}L. Marinatto, T. Weber, \textit{A Quantum Approach to Static Games of Complete Information}, Physics Letters A. 272 (2000) 291.
\bibitem{cheon 2006}T. Cheon,  I. Tsutsui, \textit{Classical and Quantum Contents of Solvable Game Theory on Hilbert Space}, Physics Letters A. 348 (2006) 147.
\bibitem{fliteny 2007}A. P. Fliteny,  L. C. L. Hollenberg, \textit{Nash Equilibria in Quantum Games with Generalized Two-Parameter Strategie}s, Physics Letters A. 363 (2007) 381.
\bibitem{makowski 2009} M. Makowski, \textit{Transitivity vs. Intransitivity in Decision Making Process - An Example in Quantum Game Theory}, Physics Letters A. 373 (2009) 2125.
\bibitem{landsburg 2011}S. Landsburg, \textit{Nash Equilibria in Quantum Games}, Proceedings of the American Mathematical Society 139 (2011) 4423.
\bibitem{arfi 2007} B. Arfi,\textit{ Quantum Social Game Theory}, Physica A 374 (2007) 794.
\bibitem{deng 2016} X. Deng, Y. Deng, Q. Liu and,  Zh.Wang, \textit{Quantum Games of Opinion Formation Based on the Marinatto-Weber Quantum Game Schem}e, Europhysics Letters 114 (2016) 50012.
\bibitem{frackiewicz 2015} P. Frackiewicz, \textit{A New Quantum Scheme for Normal-Form Games}, Quantum Information Processing  14 (2015) 1809.
\bibitem{MS2000} R. N. Mantegna and H. E. Stanley,\textit{ An Introduction
to Econophysics}, Cambridge University Press, 2000.
\bibitem{kydland 1977}F. E. Kydland, E. C. Prescott,\textit{ Rules Rather than Discretion: the Inconsistency of Optimal Plans}, Political Economy 85 (1977) 473.
\bibitem{bg 1983}R. J. Barro, D. B. Gordon, \textit{Rules, Discretion and Reputation in a Model of Monetary Policy}, Monetary Economics  12 (1983)  101.
\bibitem {lucas} R. E. Lucas and N. L. Stokey,\textit{ Optimal Fiscal and Monetary Policy in an Economy Without Capital},  Monetary Economics  12 (1983) 55.
\bibitem {albanesi} S. Albanesi, V. V. Chari, and L. Christiano, \textit{Expectation Traps and Monetary Policy}, Review of Economic Studies  70 (2003) 715.
\bibitem{king} R. G. King and R. L. Wolman, \textit{Monetary Discretion, Pricing Complementarity and Dynamic Multiple Equilibria}, Quarterly Journal of Economics, 199 (2004) 1513.
\bibitem {demirel} U. D. Demirel, \textit{The Value of Monetary Policy Commitment Under Imperfect Fiscal Credibility}, Economic Dynamics and Control 36 (2012) 813.
 \bibitem {adam} K. Adam and R. M. Billi, \textit{Distortionary Fiscal Policy and Monetary Policy Goals}, Economics Letters 122 (2014) 1.
 \bibitem{bruza} P. D. Bruza, Zh. Wang and J. R. Busemeyer, Quantum Cognition: \textit{A New Theoretical Approach to Psychology}, Feature Review 19 (2015) 383.
 \bibitem {Bordley} R.F.  Bordley,\textit{ Quantum Mechanical and Human Violations of Compound Probability Principles: Toward a Generalized Heisenberg Uncertainty Principle}, Operations Research  46 (1998) 923.
 \bibitem{Bleh} D. Bleh, T. Calarco, S. Montangero,\textit{ Quantum game of life}, Euro Physics Letters 97 (2012) 20012.
 \bibitem{Bagarello} F. Bagarello, R. Di. Salvo, F. Gargano and F. Oliveri, \textit{(H; $ \rho  $)-Induced Dynamics and the Quantum Game of Life}, Applied Mathematical Modelling 43(2017)15.
 \bibitem{Bagarello1} F. Bagarello, E. Haven,\textit{ First Results on Applying a Non-Linear Effect Formalism to Alliances between Political Parties and Buy and Sell Dynamics}, Physica A 444 (2016) 403.
 \bibitem{Haven} E. Haven, A. Khrennikov, \textit{Statistical and Subjective Interpretations of Probability in Quantum- Like Models of Cognition and Decision Making}, Journal of Mathematical Psychology 74 (2016) 82.
\bibitem {Bagarello2} F. Bagarello, F. Gargano and F. Oliveri,\textit{ A Phenomenological Operator Description of Dynamics of Crowds: Escape Strategies} , Applied Mathematical Modelling 39 (2015) 2276.
 \bibitem{Bagarello3} F. Bagarello, \textit{A Quantum-Like View to a Generalized Two Players Game}, International Journal of Theoretical Physics, 54 (2015) 3.
 \bibitem{ni} M. A. Nielsen and I. L. Chuang, \textit{Quantum Computation and Quantum Information}, Cambridge University Press, 2010.
\bibitem{strotz 1955} R. H. Strotz, \textit{Myopia and Inconsistency in Dynamic Utility Maximization}, Review of Economic Studies 23 (1955-1956)  165.
\bibitem{backus 1985} D. Backus, J. Driffill,\textit{ Inflation and Reputation}, The American Economic Review 75 (1985) 530.
\bibitem{storger 2007}J. Storger,\textit{ The Time Consistency Problem, Monetary Policy Models}, Seminar on Dynamic Fiscal Policy, Mannheim University (2007).
\bibitem{epjb2016} X. Deng, Y. Deng, Q. Liu, S. Chang, and Z.Wang, \textit{A Quantum Extension to Inspection Game}, Eur. Phys. J. B 89 (2016) 162.
\bibitem{nature2000} J. R. Friedman, V. Patel, W. Chen, S. K. Tolpygo, J. E. Lukens, \textit{Quantum Superposition of Distinct Macroscopic States}, Nature 46 (2000) 43.




\end{thebibliography}
\end{document}